\documentclass[twocolumn,showpacs,pra,amsmath,amssymb]{revtex4}% Physical Review B
\usepackage{epsfig}
\usepackage{epsfig,amsmath}
\usepackage{graphicx}% Include figure files
\usepackage{dcolumn}% Align table columns on decimal point

\newcommand{\beq}{\begin{equation}}
\newcommand{\eeq}{\end{equation}}
\newcommand{\beqa}{\begin{eqnarray}}
\newcommand{\eeqa}{\end{eqnarray}} \newcommand{\lam}{\lambda}
 \newcommand{\rh}{\rho}

\newcommand{\al}{\alpha} \newcommand{\si}{\sigma}

\def\annph#1{{ Ann.\ Phys.} {\bf #1}}
\def\josab#1{{ J. Opt.\ Soc.\ Am.\ B\/} {\bf#1}}

\def\oc#1{{ Opt.\ Commun.} {\bf#1}}

\def\pla#1{{ Phys.\ Lett. A\/} {\bf#1}}
\def\pra#1{{ Phys.\ Rev. A\/} {\bf#1}}
\def\prb#1{{ Phys.\ Rev. B\/} {\bf#1}}
\def\pre#1{{ Phys.\ Rev. E\/} {\bf#1}}
\def\prl#1{{ Phys.\ Rev.\ Lett.} {\bf#1}}
\def\sci#1{{ Science} {\bf#1}}

\begin{document}

\title{Entanglement Evolution in a Non-Markovian Environment}

\author{Ting Yu$^1$\footnote{Email address:
ting.yu@stevens.edu} and J.~H. Eberly$^2$\footnote{Email address:
eberly@pas.rochester.edu}} \affiliation{$^1$Center of Controlled
Quantum Systems and Department of Physics and Engineering Physics,
Stevens Institute of Technology,
Hoboken, New Jersey 07030-5991, USA\\
$^2$Rochester Theory Center
and Department of Physics and Astronomy, University of Rochester,
Rochester, New York 14627-0171, USA}
\date{\today}

\begin{abstract}
We extend recent theoretical studies of entanglement dynamics in the presence of environmental noise, following the long-time interest of Krzysztof Wodkiewicz in the effects of stochastic models of noise on quantum optical coherences. We investigate the quantum entanglement dynamics of two spins in the presence of classical Ornstein-Uhlenbeck noise, obtaining exact solutions for evolution dynamics. We consider how entanglement can be affected by non-Markovian noise, and discuss several limiting cases. \\

{\bf Key words}:  {Entanglement dynamics, Stochastic Schr\"odinger
Equation, Kraus Operators, Master Equations}

\end{abstract}
\pacs{03.65.Yz, 03.65.Ud, 03.67.-a}

\maketitle

\section{Introduction}

\begin{figure}[!b]
\includegraphics[width = 4cm]{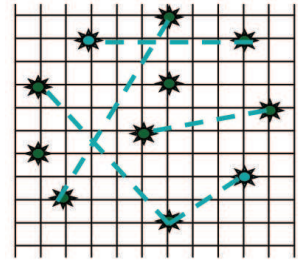}
\caption{ \label{MemNet} Sketch of remote qubits in a quantum memory net, where the dashed lines indicate entanglement, but not interaction. }
\end{figure}

A quantum system of interest may be large or small, but its background, called an environment, is almost always complex, and is often represented by a bath of bosons or fermions, or by classical random fields.  In all these cases, system dynamics is described by a quantum master equation that governs the evolution of the reduced density matrix of the system.  In the current decade, decoherence dynamics of entangled quantum systems under the influence of environmental noises has been extensively discussed in different contexts involving atoms, ions, photons, quantum dots, and Josephson junctions, to name several. This is all related to new regimes of information processing, such as quantum cryptography and quantum computation \cite{Polishfamily}. An important category of such research has treated the fascinating domain where entanglement of qubits evolves even though the qubits do not interact, even indirectly. An example is sketched in Fig. \ref{MemNet} and we restrict our attention here to this category.

In experimental environments an entangled system may be exposed to vacuum noise, phase noise, thermal noise, and various classical noises, as well as mixed combinations of noises. A number of idealized models have provided new insights by allowing entanglement evolution to be followed by solving the appropriate quantum master equation (see Zyczkowski, et al. \cite{Zyczkowski-etal}, Daffer, et al. \cite{Daffer-etal}, as well as \cite{Yu-EberlyPRL04, Yu-EberlyPRL06, Buchleitner, Davidovich-etal07, Eberly-YuSci07, Yu-EberlySci09}).  Most research on entanglement dynamics has been focused on ambient noises from environments that obey the Markov (no memory) assumption.  Recently, there is growing interest in the non-Markovian entanglement
dynamics for both discrete and continuous quantum systems (see \cite{Bellomo, Paz, Dajka, Zheng, Guo-etal, Hu-etal} and the overview in \cite{Yu-EberlySci09}).

In truth every environment is non-Markovian. Non-Markovian noise was a repeated theme in the research of Krzysztof Wodkiewicz \cite{Wodk-Eberly76, Wodk-ResFl, Wodk-ProbDist, Wodk-ColNoise, Wodk-etal84c, Wodk-etal84d, Wodk-etal84a, Wodk-etal84b, Wodk-Eberly85, Non-MarkComm, ColNoiseFirstPass, O-UFirstPass}, and we present our findings as a contribution to his scientific memory.

As far as we know there are no fully systematic investigations of non-Markovian noises or of their effect on the coherence dynamics of non-interacting spin systems. In particular, a perturbative theory leading to the Markov approximation is still lacking.  The purpose of this paper is to present a study of such problems in the simplest form. We will consider classical non-Markovian noises, modelling them as so-called Ornstein-Uhlenbeck processes, and derive the consequences for entanglement dynamics. This can be considered an extension of our earlier note on entanglement sudden death (ESD) under classical Markov noises \cite{Yu-EberlyOC06}.

%%%%%%%%%%%%%%%%%%%%%

\section{The Kubo-Anderson Model Extended to Two Qubits}
\label{modelequation}
%%%%%%%

We consider an entangled pair of spins both of which are subject to frequency fluctuations that are random \cite{Kubo_book}. We adopt a model for these fluctuations that treats them as caused by noisy environments described by Ornstein-Uhlenbeck processes. This well-known Gaussian noise model is non-Markovian in the general case but has a well-defined Markov limit. To focus exclusively on the effects on the entanglement of the spins as it arises from the noise, we assume the spins to be affected separately by separate environments, and not to interact with each other in any way, especially not through the noises. Thus the spins could be, for example, remote components of a quantum memory net (as in Fig. \ref{MemNet}) under steady attack by weak local noise. This compromises the preservation of their entanglement.

The Hamiltonian of the two-spin system can then be formally written
as (we set $\hbar=1$):
\beq
\label{Hamiltonian}
H_{\rm tot}(t)= \frac{ \Omega_A(t)}{2}\sigma_z^A +
\frac{\Omega_B(t)}{2}\sigma^B_z
\eeq
where $\Omega_A(t)$ and $\Omega_B(t)$ are the independent
fluctuations of the spin transition frequencies (level spacings). They have the mean value properties
\beqa
M[\Omega_i(t)]&=&0, \\
M[\Omega_i(t)\Omega_i(s)] &=&\alpha(t-s)\nonumber\\
&=& \frac{\Gamma_i\gamma}{2}e^{-\gamma|t-s|}, \quad i=A,B,
\eeqa
where $M[\cdot]$ stands for the statistical mean over the noises $\Omega_A(t)$ and $\Omega_B(t)$. Note that  $\gamma$ is the noise bandwidth, and $\gamma^{-1}=\tau_c$ defines the environment's finite correlation time of the noise. For simplicity, we will take the noise properties to be the same for A and B (e.g., $\Gamma_A = \Gamma_B \equiv \Gamma$), although independent. In the limit $\gamma \rightarrow \infty$, Ornstein-Uhlenbeck noise reduces to the well-known Markov case:
\beq
\alpha(t,s)=\Gamma\delta(t-s).
\eeq
For the total system described by the Hamiltonian
(\ref{Hamiltonian}), the stochastic Schr\"odinger equation is given by
\beq
i\frac{d}{dt}|\psi(t)\rangle =H_{\rm tot}(t)|\psi(t)\rangle.
\eeq
The explicit solution for the stochastic Schrodinger equation  can be
readily obtained in terms of a stochastic unitary operator:
\beq
  |\psi(t)\rangle =U(t,\Omega_A, \Omega_B)|\psi(0)\rangle,
\eeq
where the stochastic propagator $U(t,\Omega_A, \Omega_B)$ is given by
\beq U(t,\Omega_A, \Omega_B)=e^{-i\int^t_0 ds \left(\Omega_A(s)
\sigma_z^A  + \Omega_B(s)\sigma^B_z  \right)}.
\eeq
The reduced density matrix for spins A and B is then obtained from the statistical mean
\beq
  \rho(t)=M[|\psi(t)\rangle \langle \psi(t)|].
\eeq

The master equation for the reduced density matrix for the two-spin system in a non-Markovian regime can be readily derived from the stochastic Schr\"odinger equation \cite{Yu-etal04}:
\beqa \label{rhodot}
\frac{d \rho}{dt}&=&\frac{ G(t)}{4}(2\rho - \sigma_z^A \rho\sigma_z^A
- \sigma_z^B \rho\sigma_z^B ),
\eeqa
where
\beq
G(t) = \int^t_0 \alpha(t-s) ds = \frac{\Gamma}{2}(1-e^{-\gamma t}),
\eeq
The memory information of the environmental noises is encoded in the
time-dependent coefficient $G(t)$ where $\tau_c=1/\gamma$ characterizes the environmental memory time. 
In the Markov limit $\tau_c \to 0$ $ (\gamma \to \infty)$, when $G(t) \to \Gamma/2$, equation (\ref{rhodot}) reduces to the well-known Markov master equation in the presence of dephasing noises.

\section{Exact Solutions for Quantum Evolution}
\label{propagatorsection}

Solutions of master equations for the noisy evolution of two-spin
density matrices in terms of  the Kraus  operator-sum-representation
have been given before (see, for example, \cite{Yu-EberlyOC06, Yu-EberlyQIC07,Yu-Eberly2003}). In many
cases of physical interest,  the Kraus representation allows a
transparent analysis of entanglement decoherence without invoking the
explicit forms of the initial conditions.  In what follows,  we will
use the fact that for any two-spin initial state $\rho$ (pure or
mixed), the evolution of the reduced density matrix can be written
compactly as
\beq \label{Kraus}
\rho(t) = \sum_\mu K_\mu(t)\rho(0) K_\mu^\dag(t),
\eeq
where the Kraus operators $K_\mu$ satisfy $\sum_\mu
K_\mu^\dag K_\mu = 1$ for all $t$.

In order to derive the desired Kraus operators for the reduced
density matrix we begin by noting
that the solution for just spin A can be written:
\beq
|\psi(t)\rangle = U(\Omega_A, t)|\psi(0)\rangle
\eeq
where
\beq
U(\Omega_A,t) = \exp\left[-i F(t)\sigma_z\right]
\eeq
with the stochastic process $F(t)= \int^t_0 ds \Omega_A(s)$. Then our
first task is to express the stochastic
density operator $\rho_{\rm st}=|\psi(t)\rangle\langle \psi(t)|$ in
the Kraus-like operator representation form:
\beq
\label{stochoper}
\rho_{st}(t)=\exp\left[-iF(t)\sigma_z
\right]\rho(0)\exp\left[iF(t)\sigma_z \right]
\eeq
where $\rho(0)=|\psi(0)\rangle\langle \psi(0)|$ is the initial state
of the system, which is assumed to be independent of the
noise. The desired Kraus operators for the spin are
obtained by taking a statistical mean over the noise $\Omega_A(t)$ for qubit A and are given by
\beq  \label{k1}
E_1 = \left(\begin{array}{clcr}
p_A(t)  && 0\\
0 && 1\\
\end{array} \right),\,\,
E_2 = \left(\begin{array}{clcr}
q_A(t) && 0\\
0 &&  0\\
\end{array} \right),
\eeq
where the time-dependent Kraus matrix elements are
\beqa
q_A(t) &=&  \sqrt{1-p^2_A(t)},\quad {\rm and}\\
p_A(t) &=&  \exp{[-f(t)]},\quad {\rm with} \\
f(t) & \equiv & \int_0^t G(s) ds \nonumber \\
& = & \label{fDef}\frac{\Gamma}{2}[~t+\frac{1}{\gamma}(e^{-\gamma t} -
1)~],
\eeqa
and similar expressions for $p_B(t)$  and $q_B(t)$. The two-qubit
case given here can be easily applied to N noninteracting qubits, an
extension we reserve for later attention.

%%%%%%%%%%%%%%
Since our two spins are evolving independently, we have the following
four Kraus operators in terms of the tensor products of $E_1$ and
$E_2$:
\begin{eqnarray} \label{k1}
K_1 &=& \left(\begin{array}{clcr}
p_A && 0\\
0 && 1\\
\end{array} \right) \otimes  \left(
\begin{array}{clcr}
p_B & 0\\
0 & 1\\
\end{array} \right),\\
K_2 & = & \left(\begin{array}{clcr}
p_A && 0\\
0 && 1\\
\end{array}  \right)\otimes\left(
\begin{array}{clcr}
q_B & 0 \\
0 & 0\\
\end{array}\right),\\
                       K_3&=& \left(
\begin{array}{clcr}
q_A & 0\\
0 & 0 \\
\end{array} \right) \otimes \left(
\begin{array}{clcr}
p_B & 0\\
0 & 1\\
\end{array} \right),\\
                       K_4 &=& \left(
\begin{array}{clcr}
q_A & 0\\
0 & 0\\
\end{array} \right)\otimes \left(
\begin{array}{clcr}
q_B &  0 \\
0 & 0\\
\end{array} \right).\label{k5}
               \end{eqnarray}

\section{Non-Markovian entanglement dynamics}

\subsection{X matrix and concurrence}
We now consider entanglement dynamics of two half-integral spins (qubits) with an initial density matrix with the common X-form \cite{Yu-EberlyQIC07}:
\beq \label{mixedrho}
\rho^{AB} =
\begin{pmatrix}
\rho_{11} & 0 & 0 & \rho_{14}\\
0 &  \rho_{22} &  \rho_{23} & 0\\
0 &  \rho_{32} &  \rho_{33} & 0\\
\rho_{41} & 0 &  0 & \rho_{44}
\end{pmatrix}.
\eeq
Such X states occur in many contexts and include pure Bell states as
well as Werner mixed  states.

For two qubits, entanglement can be evaluated unambiguously via the
concurrence function \cite{Wootters}, which may be calculated explicitly from the density matrix $\rho^{AB}$. For qubits A and B we have: $C^{AB} = C(\rho^{AB}) = \max\{0, Q(t)\}$. Here $Q(t)$ is defined as
\beq
\label{definition_concurrence}
Q = \sqrt{\lam_1} - \sqrt{\lam_2} - \sqrt{\lam_3} - \sqrt{\lam_4},
\eeq
where the quantities $\lam_i$ are the (generally time-dependent)
eigenvalues in decreasing order of the following (nonlinear in
$\rho$) matrix:
\beq \zeta = \rho(\sigma^A_y\otimes
\sigma^B_y)\rho^*(\sigma^A_y\otimes
\sigma^B_y),\label{concurrence}
\eeq
where $\rh^*$ denotes the
complex conjugation of $\rh$ in the standard basis $|+,+\rangle,
|+,-\rangle, |-,+\rangle, |-,-\rangle$, and $\si_y$ is the usual Pauli
matrix expressed in the same basis.

 From the general solution (\ref{k1}-\ref{k5}), one can easily show for the initial state (\ref{mixedrho}) that one finds
\beqa
\label{definition_concurrence2}
Q(t) & = & 2\max
\left(|\rho_{32}(t)|-\sqrt{\rho_{11}(0)\rho_{44}(0)}, \right.
\nonumber \\
&&\left. |\rho_{14}(t)|-\sqrt{\rho_{22}(0)\rho_{33}(0)}\right).
\eeqa

\subsection{Solutions for non-Markovian disentanglement}

The Ornstein-Uhlenbeck phase-noise solutions for the density matrix elements of a general initial state are given by
\beqa
\rho_{12}(t)&=&\rho_{12}(0)e^{-f(t)}, \\
\rho_{13}(t)&=&\rho_{13}(0)e^{-f(t)}, \\
\rho_{24}(t) &=&\rho_{24}(0)e^{-f(t)}, \\
\rho_{34}(t) &=& \rho_{34}(0)e^{-f(t)}\\
\rho_{23}(t) &=&\rho_{23}(0)e^{-2f(t)}, \\
\rho_{14}(t) &=& \rho_{14}(0)e^{-2f(t)}\\
\rho_{ii}(t) &=& \rho_{ii}(0) \,\,\, (i=1,2,3,4)
\eeqa
where $f(t)$ is defined in (\ref{fDef}).  Let us note that in the
limit $\gamma \rightarrow \infty$, we recover  the standard Markov
approximation where $f(t) = \Gamma t/2$.

Although there is no compact analytical expression for the
concurrence $C(\rho(t))$ with an arbitrary initial state, we can
readily show that preservation of entanglement is restricted by the
inequality
\beq
\label{impact}
C(\rho(t)) \leq e^{-2f(t)}C(\rho(0)).
\eeq

A sharper result occurs for the X matrix under consideration because
the diagonal elements are independent of $t$. Since they all
vanish as $\exp(-f(t))$ for increasing
$t$, we know that $Q(t)$ must eventually become strictly negative if
diagonal values are initially non-zero (e.g., any finite-temperature equilibrium state). Negative $Q$ mandates $C^{AB} = 0$, so ESD must occur. Next we will consider key limiting cases.

\subsection{Entanglement decay:  Stationary limit}
We consider now the stationary limit $\gamma t \gg 1$. Then
\beq
f(t)=\frac{\Gamma}{2}[t+\frac{1}{\gamma}(e^{-\gamma t}-1)] \to
\frac{\Gamma}{2} t
\eeq
Therefore, from (\ref{impact}), we get
\beq
C(\rho(t)) \leq e^{- \Gamma t}C(\rho(0)).
\eeq
Hence, the entanglement decay rate is at least as rapid as $\Gamma$, and may be much faster. Clearly, the stationary limit is identical to the Markov limit $\alpha(t-s) = \Gamma \delta(t-s)$.

\subsection{Entanglement decay:  Short-time limit}
Now let us turn to the opposite limiting case: $\gamma t \ll 1$. In this case, we can use the following approximation,
\beq
e^{-\gamma t} \sim 1-\gamma t +\frac{1}{2}\gamma^2 t^2
\eeq
Therefore,
\beq
f(t)=\frac{1}{4}\Gamma \gamma t^2
\eeq
Then concurrence decay is bounded by
\beq
C(\rho(t)) \leq e^{-\frac{1}{2}\Gamma \gamma  t^2}C(\rho(0))
\eeq
Hence, the effective disentanglement time is given by:
\beq
\tau_{\rm dis}=\sqrt{\frac{{2}}{ {\Gamma \gamma}}}
\eeq

Clearly, for non-Markovian noises, the short-time limit is more
interesting since it shows that the resultant entanglement behavior
deviates significantly from the well-known Markov dynamics. Obviously, the smaller $\gamma$ is, the better approximation we have.
For both limiting cases, it is easy to prove that a sufficient
condition for ESD to occur is $\rh_{11} \rh_{22} \rh_{33} \rh_{44}
\neq 0$ \cite{HuangandZhu}.

\begin{figure}[!t]
\includegraphics[width = 6cm]{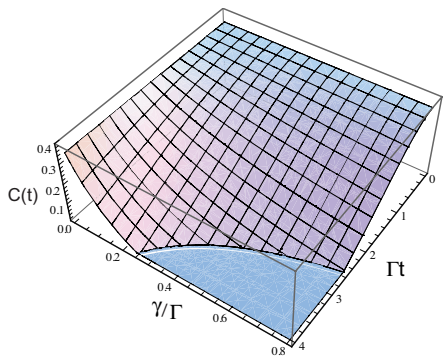}
\caption{ \label{fig.CNet} The graph shows  $C^{AB}$ vs. $\Gamma t$ and
$\gamma/\Gamma$. The value $\alpha = 1/3$ has been chosen. The reservoir bandwidth $\gamma$ controls the approach to the Markov limit, and we see that for $\gamma < \Gamma$ the inevitable onset of ESD (the region where $C(t) = 0$) can be substantially delayed. }
\end{figure}

\section{Evaluation for a special X state}
In Fig. \ref{fig.CNet} we show the evolution of the concurrence for a specific X-form entangled state:
\beq
\rho^{AB}_\alpha (0) =\frac{1}{3}
\begin{pmatrix}
\alpha  & 0 & 0 & 0\\
0 &  1 &  1 & 0\\
0 &  1 &   1 & 0\\
0  & 0 &  0 & 1-\alpha
\end{pmatrix},
\eeq
where $0\leq \alpha \leq 1$, so the initial concurrence is $C(0) = 2/3[1-\sqrt{\al(1-\al)}] > 0$.

The time dependence of the concurrence of this state is well-known in the Markov dephasing limit \cite{Yu-EberlyOC06,Yu-EberlySci09}. For our present non-Markovian case, which introduces the environmental bandwidth $\gamma$, the time dependence of the $Q$ parameter satisfies
\beq
Q(t) = \frac{2}{3}\Big(e^{-f(t)} - \sqrt{\alpha(1-\alpha)}\Big).
\eeq
For all finite values of $\gamma$ the quantity $e^{-f(t)}$ approaches
zero exponentially at long times, and then $Q(t)$ must become negative, so ESD inevitably occurs, with a finite disentanglement time $t_{\rm ESD}$ given by
\beq
e^{-2f(t_{\rm ESD})} = \al(1-\al).
\eeq

\begin{figure}[!t]
\includegraphics[width = 6cm]{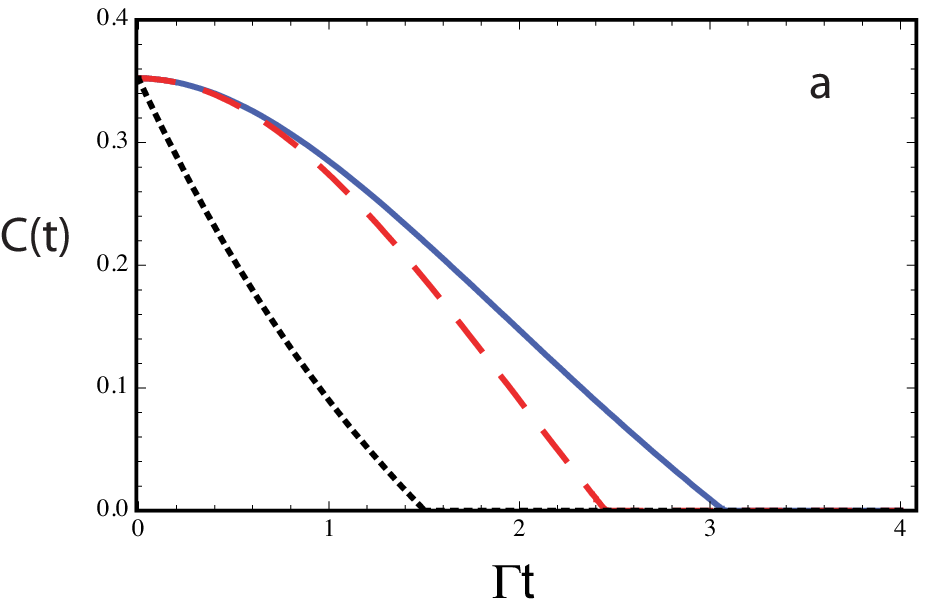}
\includegraphics[width = 6cm]{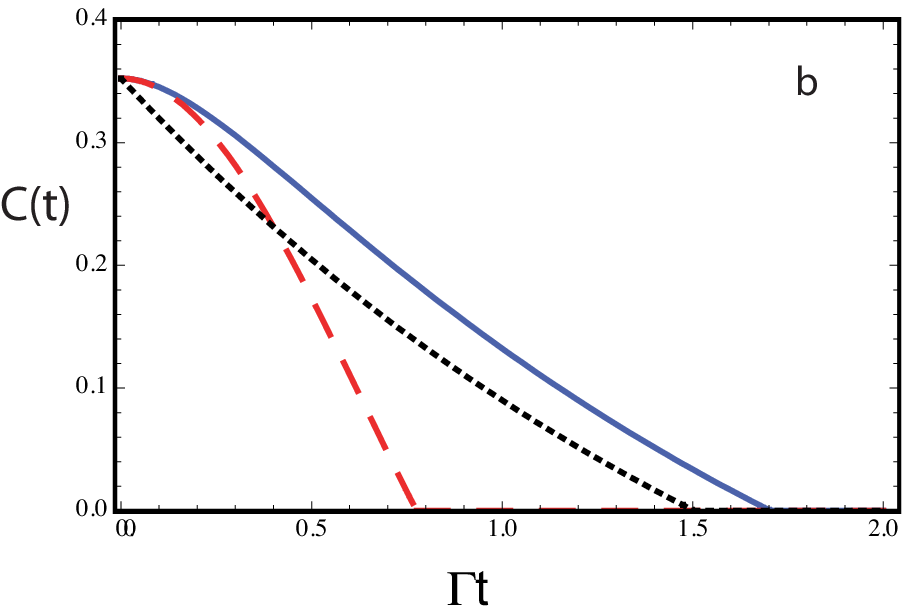}
\caption{ \label{special} The graphs show  $C_\alpha^{AB}$ vs. $t$  with
$\alpha=1/3$.    Fig. 3a shows that with $\gamma/\Gamma=0.5$, an initially non-Markovian entanglement (solid line) evolves in a markedly different way compared to its Markov limit (dotted line). Clearly, the short-time limit (dashed line) gives a better approximation than the stationary limit (dotted line).  However, as shown in Fig.3b where $\gamma/\Gamma=5$, the difference is washed out at later times when the short-time limit (dashed line) ceases to be a good approximation.}
\end{figure}

Fig. \ref{special} shows several interesting features of entanglement evolution under Ornstein-Uhlenbeck noise. Clearly, we see that non-Markovian noises have markedly different affects on entanglement evolution at short times, while the long time limit gives rise to familiar Markov behavior. First, note that ESD must occur for the entire parameter range of $\alpha$, except for the end points $\alpha=0,1$, given our initial X-state. However, the ESD times are different for non-Markovian short-time and stationary limit cases. From Fig.~3a, it should be noted that the disentanglement time for non-Markovian regimes could be significantly longer than the disentanglement times in the  Markov limit if the dissipation is small. However, once the state becomes separable it will never become entangled again. That is, entanglement rebirth or revival does not occur for Ornstein-Uhlenbeck noise \cite{Guo-etal}.

\section{Concluding remarks}

We have presented here, as a contribution to the scientific memory of Krzystof Wodkiewicz, the first results of a new investigation that has clear connections to his long-time interest in quantum systems evolving under the influence of stochastic perturbations. One of the targets of his creativity and energy, over many years, was the challenge presented by the influence of non-Markovian noise and we have addressed that challenge with some calculations focused on entanglement.

We have shown that entanglement dynamics under Ornstein-Uhlenbeck noise can be affected in several different ways, depending on the initial entangled states and the noise correlation time. It can be seen that the non-Markovian properties can prolong the life of entanglement. We note that the effective long-time relaxation rate $\Gamma$ is ordinarily associated with experimentally accessible relaxation times such as $T_1$ and $T_2$. Fig. \ref{fig.CNet} highlights the unusual domain $\gamma < \Gamma$, in which these times are shorter than the internal environmental relaxation time $\tau_c$. Entanglement survival is of fundamental interest at short times in quantum information processing (see \cite{Gordon, Zhang-etal}). In the case of the  short-time limit our results capture the features of quadratic rather than exponential decay at early times.  In this simple model, non-Markovian noises appear to  play a role as a short-time decoherence buffer,  but entanglement measured by concurrence will inevitably conform to the stationary limit at long times.  Finally, it is interesting to note that our findings based on the classical phase noise model can be extended into the case of quantum phase noises where the environment is modeled as a set of harmonic oscillators at a finite 
temperature \cite{Yu-EberlyPRB02}.

\section*{Acknowledgements}
TY and JHE  acknowledge partial financial support from the
following agencies: DARPA  HR0011-09-1-0008, ARO 48422-PH, NSF PHY-0601804, NSF PHY-0925174.


\begin{thebibliography}{99}

\bibitem{Polishfamily} R. Horodecki, P. Horodecki, M. Horodecki  and K. Horodecki, Rev. Mod. Phys. {\bf 81}, 865(2009).

\bibitem{Zyczkowski-etal} K. Zyczkowski, et al. \pra{65}, 012101
(2001).

\bibitem{Daffer-etal} S. Daffer, K. Wodkiewicz and J. K. McIver,
\pra {67}, 062312 (2003).

\bibitem{Yu-EberlyPRL04} T. Yu and J. H. Eberly, \prl {93}, 140404
(2004).

\bibitem{Yu-EberlyPRL06} T. Yu and J. H. Eberly, \prl {97}, 140403
(2006).

\bibitem{Buchleitner} F. Mintert, A. R. R. Carvalho, M.\ Kus, and A. Buchleitner, Phys. Rep.  {\bf 415}, 207 (2005).

\bibitem{Davidovich-etal07} M. P. Almeida et al, Science  {\bf 316}, 579 (2007).

\bibitem{Eberly-YuSci07} J. H. Eberly and T. Yu, \sci{316}, 555 (2007).

\bibitem{Yu-EberlySci09} T. Yu and J.H. Eberly, \sci {323}, 598 (2009), and references therein.

\bibitem{Bellomo} B. Bellomo, R. Lo Franco, G. Compagno, \prl{99}, 160502 (2007).

\bibitem{Paz} J. P. Paz and A. J.  Roncaglia, \prl{100}, 220401 ( 2008).

\bibitem{Dajka} J. Dajka, M. Mierzejewski and J. Luczka, \pra{77}, 042316 (2008).

\bibitem{Zheng} X. F. Cao and H. Zheng, \pra{77}, 022320 (2008).

\bibitem{Guo-etal}  For recent experimental research involving non-Markovian entanglement dynamics, see J. S. Xu et al, arXiv:0903.5233.

\bibitem{Hu-etal} C. Anastopoulos, S. Shresta, and B. L. Hu, arXiv: quant-ph/0610007.

\bibitem{Wodk-Eberly76} K. Wodkiewicz and J.H. Eberly,
Markovian and non-Markovian behavior in 2-level atom fluorescence,
\annph {101}, 574-593 (1976).

\bibitem{Wodk-ResFl} K. Wodkiewicz, Non-Markovian resonance
fluorescence, \pla {73}, 94-96 (1979).

\bibitem{Wodk-ProbDist} K. Wodkiewicz, Functional representation of a
non-Markovian probability-distribution in statistical-mechanics, \pla
{84}, 56-58 (1981).

\bibitem{Wodk-ColNoise} K. Wodkiewicz, Langevin-equations with
colored noise in quantum optics, Acta Phys. Pol. A  {\bf 63}, 191-200
(1983).

\bibitem{Wodk-etal84a} J. H. Eberly, K. Wodkiewicz and B. W. Shore,
Noise in strong laser-atom interactions - phase telegraph noise, \pra
{30}, 2381-2389 (1984).

\bibitem{Wodk-etal84b} K. Wodkiewicz, B. W. Shore and J. H. Eberly,
Noise in strong laser-atom interactions - frequency fluctuations and
nonexponential correlations, \pra {30}, 2390-2398 (1984).

\bibitem{Wodk-etal84c} K. Wodkiewicz, B. W. Shore and J. H. Eberly,
Pre-Gaussian Noise In Strong Laser Atom Interactions, \josab {1},
398-405, (1984).

\bibitem{Wodk-etal84d} K. Wodkiewicz, J. H. Eberly and B. W. Shore,
Phase and frequency jump theory of laser band shape, \josa {1},
506-506   (1984).

\bibitem{Wodk-Eberly85} K. Wodkiewicz and J. H. Eberly,
Shot noise and general jump-processes in strong laser-atom
interactions, \pra {31}, 2314-2317 (1985).

\bibitem{Non-MarkComm} K. Wodkiewicz, Spontaneous and induced
emission of soft bosons - Exact non-Markovian solution - Comment,
\prl {63},   2693-2693 (1989).

\bibitem{ColNoiseFirstPass} M. Kus,  E. Wajnryb and K. Wodkiewicz,
Mean 1st-passage time in the presence of colored noise - a
random-telegraph-signal approach, \pra {43}, 4167-4174 (1991).

\bibitem{O-UFirstPass} M. Kus and K. Wodkiewicz, Mean 1st-passage
time in the presence of telegraph noise and the Ornstein-Uhlenbeck
process, \pre {47}, 4055-4063 (1993).

\bibitem{Yu-EberlyOC06} T. Yu and J. H. Eberly, \oc {264}, 393 (2006).

\bibitem{Kubo_book}  R. Kubo, M. Toda, and N. Hashitsume,  {\it
Statistical Physics II}, (Berlin, Springer, 1991).

\bibitem{Yu-etal04} W. Strunz and T. Yu, \pra {69}, 052115 (2004); and T.  Yu, \pra {69},   062107 (2004).

\bibitem{Yu-EberlyQIC07} T. Yu and J. H. Eberly, Quant. Inf. and Comp.
{\bf 7}, 459 (2007).

\bibitem{Yu-Eberly2003} T. Yu and J. H. Eberly, \prb {68}, 165322  (2003).

\bibitem{Wootters} W. K. Wootters, \prl {80}, 2245 (1998).

\bibitem{HuangandZhu} J. H. Huang and S. Y. Zhu, \pra{76}, 062322 (2007) and \oc{281}, 2156 (2008).

\bibitem{Gordon} D. Gordon, Euro. Phys. Lett. {\bf 83}, 30009 (2008).

\bibitem{Zhang-etal} J. H. An, Y. Yeo and W. M. Zhang, J. Phys. A {\bf 42}, 015302  (2009).

\bibitem{Yu-EberlyPRB02} T. Yu and J. H. Eberly,  \prb{66},  193306 (2002).

\end{thebibliography}
\end{document}